\documentstyle [12pt] {article}
\input epsf

\newcommand{\beq}{\begin{equation}}
\newcommand{\eeq}{\end{equation}}
\newcommand{\beqs}{\begin{eqnarray}}
\newcommand{\eeqs}{\end{eqnarray}}
\newcommand{\gtwid}{\mathrel{\raise.3ex\hbox{$>$\kern-.75em\lower1ex
\hbox{$\sim$}}}}
\newcommand{\ltwid}{\mathrel{\raise.3ex\hbox{$<$\kern-.75em\lower1ex
\hbox{$\sim$}}}}

\begin{document}

\def\thefootnote{\fnsymbol{footnote}}
\baselineskip 6.0mm

\begin{flushright}
\begin{tabular}{l}
ITP-SB-96-43 
\end{tabular}
\end{flushright}

\vspace{4mm}
\begin{center}

{\bf Complex-Temperature Phase Diagrams of }  

\vspace{4mm}

{\bf 1D Spin Models with Next-Nearest-Neighbor Couplings}

\vspace{12mm}

\setcounter{footnote}{0}
Robert Shrock\footnote{email: shrock@insti.physics.sunysb.edu}
\setcounter{footnote}{6}
and Shan-Ho Tsai\footnote{email: tsai@insti.physics.sunysb.edu}

\vspace{6mm}
Institute for Theoretical Physics  \\
State University of New York       \\
Stony Brook, N. Y. 11794-3840  \\

\vspace{12mm}

{\bf Abstract}
\end{center}

   We study the dependence of complex-temperature phase diagrams on details of
the Hamiltonian, focusing on the effect of non-nearest-neighbor spin-spin 
couplings.  For this purpose, we consider a simple exactly solvable model, 
the 1D Ising model with nearest-neighbor (NN) and next-to-nearest-neighbor 
(NNN) couplings. We work out the exact phase diagrams for various values of 
$J_{nnn}/J_{nn}$ and compare these with the case of pure nearest-neighbor 
(NN) couplings.  We also give some similar results for the 1D Potts model 
with NN and NNN couplings. 

\vspace{2mm}

\begin{flushleft}
PACS nos. 05.50.+q, 05.70.Fh, 64.60.-i, 64.60.Cn
\end{flushleft}

\vspace{16mm}

\pagestyle{empty}
\newpage

\pagestyle{plain}
\pagenumbering{arabic}
\renewcommand{\thefootnote}{\arabic{footnote}}
\setcounter{footnote}{0}

\section{Introduction} 
\label{intro}

   Yang and Lee pioneered a very interesting line of research in which one
studies statistical mechanical models with the external magnetic field $H$ 
generalized from real to complex values \cite{yl}. Since the free energy of
a spin system $f=f(K,H)$ is a function of both the field $H$ and the 
temperature $T$ (or equivalently, $K= \beta J$, where $\beta = (k_BT)^{-1}$ 
and $J$ is the spin-spin coupling), a related complexification is to 
generalize $K$ from real to complex values. Both of these complexifications 
give one deeper insight into the properties of such models. Recently, there has
been renewed interest in this subject for Ising models \cite{ms}-\cite{gehs} 
and Potts models \cite{mbook}-\cite{pfef} (earlier references can be found
in these papers).  As the comparison of complex-temperature phase diagrams 
for the 2D (isotropic, nearest-neighbor) Ising model with spin $s \ge 1$ 
versus $s=1/2$ in Refs. \cite{egj}, 
\cite{hs}-\cite{gehs} has shown, these phase diagrams differ considerably for
different $s$ even though all values of $s$ are in the same universality
class for the usual paramagnetic (PM) to ferromagnetic (FM) phase transition 
in this model.  

  Another parameter of the Hamiltonian on which complex-temperature phase 
diagrams depend is the ratio of spin-spin couplings. Indeed, in cases such as 
the (spin 1/2, nearest-neighbor) Ising model on regular 
bipartite 2D lattices, where a variation in the ratio of spin-spin couplings 
along different lattice directions does not change the universality class of 
the PM to FM phase transition, this variation has a significant 
effect on the continuous locus of points where the free energy is non-analytic,
which is denoted ${\cal B}$: while ${\cal B}$ is a one-dimensional algebraic
variety for the case of isotropic couplings, it becomes a two-dimensional
variety for the non-isotropic case \cite{aniso}.  Recently, we have also 
shown, using exact results, that at complex-temperature
singularities, the exponents describing the behavior of various thermodynamic
functions depend, in general, on lattice type, which constitutes a 
violation of universality \cite{chisq}.  Moreover, we have found a number of 
violations of exponent relations at such singularities, such as 
$\gamma \ne \gamma'$ (for the susceptibility exponent) and 
$\alpha + 2\beta + \gamma \ne 2$ (as one approaches such a singularity from
the PM phase) \cite{chisq}.  

    In the present paper, we explore further the extent of non-universal
features of complex-temperature phase diagram of spin models, focussing 
on the effect of non-nearest-neighbor spin-spin couplings.  Here, by
``non-universal'', we mean features of the complex-temperature phase diagram
that depend on a parameter in the Hamiltonian, where a variation of this
parameter does not change the universality class of a phase transition at a
given critical point.  
For our study, it will suffice to use a simple exactly solvable
model, namely, the 1D Ising model with nearest-neighbor (NN) and
next-to-nearest-neighbor (NNN) spin-spin couplings, $J_{nn}$ and $J_{nnn}$.
This model (except for the special case $J_{nnn}=-|J_{nn}|/2$) is critical at
$T=0$, so we specifically study the dependence of the complex-temperature
phase diagram on the ratio 
\beq
r = \frac{J_{nnn}}{J_{nn}}
\label{r}
\eeq
for the range of $r$ values where changes in $r$ do not change the ground state
of the model or the universality class of the critical point at $T=0$. After
some early papers \cite{m,oguchi}, a detailed solution and discussion of this 
model was given by Stephenson in Ref. \cite{steph70} and subsequent papers
\cite{steph2}.  Some later papers include Refs. \cite{selke,redner}. 
All of these dealt with physical temperature; the model has not, to our
knowledge, been studied for complex temperature.  
As our analysis of the complex-temperature 
phase diagram of this model will demonstrate, it illustrates very well 
how sensitive the CT phase diagram is to the presence of such NNN 
couplings. This example is also useful in giving one a 
qualitative idea, in a simple context, of what to expect
concerning the effect of non-nearest-neighbor couplings in higher-dimensional
spin models for which one does not have any exact solution.  For these 
higher-dimensional models the addition of NNN couplings gives rise to
quite complicated phase diagrams even for physical temperature \cite{annni}. 
Moreover, although $d=1$ is
the lower critical dimensionality for the Ising model, and some features, such
as the lack of a physical phase transition at finite temperature are
qualitatively different from the behavior in $d > d_{\ell.}=1$, past experience
with $2+\epsilon$ and $1+\epsilon$ expansions \cite{eps} has shown that one can
learn about spin models in intermediate dimensionalities by moving upward from 
$d_{\ell}$ as well as downward from the upper critical dimensionality $d_{u}$.
Indeed, several intriguing connections were previously noted between the CT 
phase diagrams of the 1D spin $s$ Ising model and features of the same model 
on the square lattice \cite{hs,is1d}. In passing, it may be noted that the 1D 
NNN Ising model has also been used in a complementary manner in Ref. \cite{ko},
for a study of Yang-Lee (complex-field) zeros of the partition function for 
physical temperature.  

\section{Ising Model and Notation}
\label{model}

   In this section, we shall briefly give the relevant notation and review 
some basic properties of the model for physical temperature which will serve as
a background for our new results on the complex-temperature properties. 
The 1D (spin 1/2) Ising model with NN and NNN couplings 
is defined, for temperature $T$ and external
magnetic field $H$ on a 1D lattice, by the partition function
$Z = \sum_{\{ \sigma_n \}} e^{-\beta {\cal H}}$ with Hamiltonian 
\beq
{\cal H} = -J_{nn}\sum_n \sigma_n \sigma_{n+1} - J_{nnn}\sum_{n} \sigma_n
\sigma_{n+2} - H \sum_n \sigma_n
\label{ham}
\eeq
where $\sigma_n = \pm 1$ and $\beta = (k_BT)^{-1}$.  Except where otherwise
indicated, we take $H=0$ below.  It is convenient to define 
\beq
K = \beta J_{nn}
\label{k}
\eeq
\beq
K' = \beta J_{nnn}
\label{kp}
\eeq
and $h = \beta H$. 
Recall that on a bipartite lattice $\Lambda$, 
without loss of generality, one may take
$J_{nn} \ge 0$ (if $J_{nn} < 0$ initially, we can redefine 
$\sigma_n \to -\sigma_n$ for $n \in \Lambda_e$, $\sigma_n \to \sigma_n$
for $n \in \Lambda_o$, and $J_{nn} \to -J_{nn}$ where $\Lambda_e$ and
$\Lambda_o$ denote the even and odd sublattices of $\Lambda$; the partition
function $Z$ is invariant under this mapping).  Using this fact, we shall thus
take $J_{nn} > 0$ henceforth.  We define the ratio of couplings by
eq. (\ref{r}). 
We shall mainly consider the effect of NNN couplings with $r \ge 0$ since 
(given that we take $J_{nn} > 0$), these NNN couplings do not introduce any
frustration or competition and do not change the universality class or ground
state of the model.  In contrast, given that we take $J_{nn} > 0$, a NNN
coupling with negative $J_{nnn}$ does introduce such competition
and frustration and is not necessarily an irrelevant perturbation
to the Hamiltonian. We shall also include some results for negative $J_{nnn}$. 
It is convenient to define the Boltzmann weight variables $z=e^{-2K'}$, 
\beq
u = z^2 = e^{-4K'}
\label{udef}
\eeq
and $z_{_K}=e^{-2K}$, with 
\beq
u_{_K} = z_{_K}^2 = e^{-4K}
\label{ukdef}
\eeq
For our study, it will be sufficient to consider the cases
where (i) $r=1/p$ where $p$ is an integer, or (ii) $r$ is an
integer, since these already amply demonstrate the sensitivity of the
complex-temperature phase diagram to the value of $r$. 
In these cases, $Z$ is a generalized polynomial (i.e. with 
positive and negative integer powers) in the respective Boltzmann weights (i)
$u$ and (ii) $u_{_K}$.  Of course, one could also consider the case $r=p/q$
where $p \ne 1$ and $q \ne 1$ are relatively prime integers, but we shall not 
do this here.  The (reduced, per site) free energy is defined as 
$f = -\beta F = \lim_{N \to \infty} N^{-1} \ln Z$ in the thermodynamic
limit.  We assume periodic boundary conditions and take the number of lattice
sites $N$ to be even in order to preserve the bipartite lattice structure on a
finite lattice. 

\section{Generalities and Complex-Temperature Phases}

   For $d=1$ dimension, it is straightforward to solve this model exactly, 
e.g., by transfer matrix methods. One has 
\beq
Z = Tr({\cal T}^N) = \sum_j \lambda_j^N
\label{zeq}
\eeq
where the $\lambda_j$, $j=1,...4$ denote the eigenvalues of the
transfer matrix ${\cal T}$ defined by
${\cal T}_{nn'}= \langle \ell_n|\exp(-\beta E(\ell_n, \ell_{n'}))|\ell_{n'}
\rangle$.  It is natural to analyze the phase diagram in complex plane of the 
appropriate Boltzmann weight variables (such as $u$ or $u_{_K}$ for positive
$r$).  For physical temperature, phase transitions 
are associated with degeneracy of leading eigenvalues. 
There is an obvious generalization of this to the case of complex
temperature: in a given region of $u$ or other Boltzmann weight variable, 
the eigenvalue of ${\cal T}$
which has maximal magnitude, $\lambda_{max}$  gives the
dominant contribution to $Z$ and hence, in the thermodynamic
limit, $f$ receives a contribution only from $\lambda_{max}$:
$f=\ln ( \lambda_{max})$.  For complex $K$, $f$ is, in general,
also complex.  The CT phase boundaries are determined by the degeneracy, in
magnitude, of leading eigenvalues of ${\cal T}$.
As one moves from a region with one dominant eigenvalue
$\lambda_{max}$ to a region in which a different eigenvalue $\lambda_{max}'$
dominates, there is a non-analyticity in $f$ as it switches from
$f=\ln (\lambda_{max})$ to $f=\ln (\lambda_{max}')$.  The boundaries of these
regions are defined by the degeneracy condition among dominant eigenvalues,
$|\lambda_{max}|=|\lambda_{max}'|$.  These form curves in the plane of the
given Boltzmann weight variable.  

   Of course, for physical temperature, 
a 1D spin model with finite-range interactions has no
non-analyticities for any (finite) value of $K$, so that, in particular, the 1D
NNN Ising model is analytic along the positive real axis in the complex $u$ 
or $u_{_K}$ plane and is only singular at $T=0$.  In this context, we recall
that the elements of the transfer matrix are non-negative (positive or zero)
real functions of $T$ for physical temperature, and the Perron-Frobenius
theorem \cite{bellman} guarantees that a (finite-dimensional, but
not necessarily symmetric) square matrix with non-negative real entries has a
real positive eigenvalue of greatest magnitude. This property 
underlies the absence of any non-analyticity and associated phase transition in
a 1D spin model with finite-range interactions.  However, when one
generalizes the temperature to complex values, the elements of the transfer
matrix are not, in general, non-negative (they are complex), so that the 
premise of the Perron-Frobenius
theorem is no longer satisfied, and, indeed, the maximal 
eigenvalue can switch as one varies $K$ over complex values. 

  Since the $\lambda_j$ are analytic functions of $u$, whence
$\lambda_j(u^*)=\lambda_j(u)^*$, it follows that the solutions to
the degeneracy equations defining the boundaries between different phases,
$|\lambda_i|=|\lambda_j|$, are invariant under $u \to u^*$.  Hence the
complex-temperature phase boundary ${\cal B}$, or equivalently, the continuous
locus of points where the free energy $f$ is non-analytic, 
is invariant under $u \to u^*$. The same applies for ${\cal B}$ in the 
$u_{_K}$ plane, as discussed below. 
Although the model has a physical phase structure consisting only of the
$Z_2$-symmetric, disordered phase, its complex-temperature phase diagram 
is nontrivial and exhibits a number of interesting features. 

   Since the model has NNN couplings, one cannot define the transfer matrix as
acting just between states consisting only of neighboring spins.  The most
compact way to define the transfer matrix is to use state vectors consisting of
pairs of spins: $v_n=|\sigma_n, \sigma_{n+1}\rangle$. Then 
\beq 
\langle v_n|{\cal T}|v_{n+1} \rangle = 
\langle v_n |e^{-\beta {\cal H}}|v_{n+1}\rangle = 
\exp \Bigl ( \frac{K}{2}(\sigma_n\sigma_{n+1} + \sigma_{n+1}\sigma_{n+2} ) 
+ K'\sigma_n\sigma_{n+2} + h\sigma_{n+1} \Bigr ) 
\label{t}
\eeq
The factor of (1/2) is included because each interaction of spins within a 
given vector $|v_n \rangle$ is counted twice in the sum over $n$.  Hence, with
the basis vectors ordered as 
$\{|v_n \rangle = |++ \rangle, |+- \rangle, |-+ \rangle, |-- \rangle \}$, 
one has \cite{m,oguchi} 
\beq
{\cal T} =  \left (\begin{array}{cccc}
  e^{K+K'+h}  & e^{-K'+h}    & 0             & 0           \\
  0           & 0            & e^{-K+K'-h}   & e^{-K'-h}   \\
  e^{-K'+h}   & e^{-K+K'+h}  & 0             & 0           \\
  0           & 0            & e^{-K'-h}     & e^{K+K'-h} \end{array}\right )
\label{tmatrix}
\eeq
Note that ${\cal T}$ has zero matrix elements if the second spin in 
$|v_n \rangle$ has
a value different from the the first spin in $|v_{n+1}\rangle$, since these
states overlap in this middle spin.  Although ${\cal T}$ is not symmetric, the 
usual relation $Z = Tr({\cal T}^N) = \sum_j \lambda_j^N$, where the 
$\lambda_j$'s are the eigenvalues of ${\cal T}$, still holds; this follows 
 from (i) the theorem \cite{bellman} that an arbitrary complex 
$\ell \times \ell$ matrix can be put into upper triangular (u.t.) form by a 
unitary transformation $V$: $V {\cal T} V^{-1} = {\cal T}_{u.t.}$, such that 
$diag({\cal T}_{u.t.}) = \{ \lambda_1, ... \lambda_\ell \}$; and (ii) the 
identity $Tr({\cal T}_{u.t.}^N) = Tr({\cal T}^N) = 
\sum_{j=1}^{\ell} \lambda_j^N$. It is convenient to define 
$\bar {\cal T} = e^{-(K+K'+h)}{\cal T}$ so that $\bar {\cal T}_{11}=1$ and 
consider the eigenvalues of $\bar {\cal T}$. We shall take $h=0$ henceforth. 

  $\bar {\cal T}$ has the eigenvalues 
\beq
\lambda_{1 \pm} = e^{-K}\Bigl [ \cosh K \pm \sqrt{\sinh^2 K + e^{-4K'}} \ \ 
\Bigr ]
\label{lam1pm}
\eeq
\beq
\lambda_{2 \pm} = e^{-K}\Bigl [ \sinh K \pm \sqrt{\cosh^2 K - e^{-4K'}} \ \ 
\Bigr ]
\label{lam2pm}
\eeq
For physical temperature, $\lambda_{1+}$ is the dominant eigenvalue, so the
(reduced, per site) free energy is $f=\ln(\lambda_{1+})$.  In passing, we
note that an equivalent method for solving the model in zero field is to 
re-express it formally in terms of a different theory with only NN couplings 
but a nonzero effective field \cite{steph70}. 

The internal or configurational energy (per site) $U$ is 
\beq
U = -\frac{J_{nn} \sinh(K) }{ \sqrt{\sinh^2(K) + e^{-4K'}}} 
- J_{nnn}\Biggl [ 1 - \frac{2e^{-4K'}}{\sinh^2(K) + e^{-4K'} + 
\cosh(K)\sqrt{\sinh^2(K)+e^{-4K'}}} \Biggr ] 
\label{uenergy}
\eeq
Observe that
\beq
U(J_{nn},J_{nnn},\beta) = U(-J_{nn},J_{nnn},\beta)
\label{usym}
\eeq
which is an explicit illustration of the general fact noted above that we can, 
without loss of generality, take $J_{nn} > 0$.  The nature of the
ground state (g.s.) depends on $r$ \cite{oguchi}: if $J_{nnn}$ is positive or 
sufficiently weakly negative, the ground state is ferromagnetic: 
\beq
r > -\frac{1}{2} \ \Longrightarrow \ {\rm FM \ \ g.s.} 
\label{fmregime}
\eeq
while for stronger negative $J_{nnn}$ it changes according to 
\beq
r < -\frac{1}{2} \ \Longrightarrow \ (2,2) \ \ {\rm g.s.}
\label{22regime}
\eeq
where the $(2,2)$ g.s. refers to a spin configuration of the modulated form
$(++--++--...)$   Correspondingly, there is a non-analytic change in the ground
state energy: 
\beq
U(T=0)=-(J_{nn}+J_{nnn}) \quad {\rm for} \quad r \ge -\frac{1}{2}
\label{uvalue1}
\eeq
whereas 
\beq
U(T=0)=J_{nnn} \quad {\rm for} \quad r \le -\frac{1}{2}
\label{uvalue2}
\eeq
Evidently (given that we take $J_{nn} > 0$), negative values of
$J_{nnn}$ give rise to competing interactions and frustration.  Indeed, if
$J_{nnn}$ is sufficiently negative that $r < -1/2$, it changes the ground 
state of the model. In our study of the dependence of the 
complex-temperature phase diagram on the addition of irrelevant operators 
(i.e., irrelevant in the sense that they do not change the $T=0$ critical 
behavior), we therefore shall restrict to the case $r > -1/2$. However, since
the range $r \le -1/2$ is of interest in its own right, we shall also briefly
digress to discuss this case further below.  As Stephenson showed 
\cite{steph70,steph2}, 
even in the range $-1/2 < r < 0$, where the ground state still
exhibits saturated FM long-range order,
the NNN coupling has the interesting effect of giving rise to a ``disorder
temperature'' $T_D$, where the correlation length has a local mininum; for 
$T < T_D$, the spin-spin correlation functions have a purely exponential 
asymptotic decay, while for $T > T_D$, their asymptotic decay is an exponential
multiplied by an oscillatory factor.  

   The $T=0$ criticality of the model is typical of a theory at its lower
critical dimensionality, here $d=1$. As $T \to 0$, the specific heat $C$ 
has an essential zero given by 
\beq
C \sim 4k_B(1+2r)^2 K^2 e^{-2(1+2r)K} \quad {\rm as} \ T \to 0 \quad 
{\rm for} \ \  r > -\frac{1}{2}
\label{crgtmhalf}
\eeq
and 
\beqs
C & \sim & \frac{k_B}{2}(1+2r)^2 K^2 e^{(1+2r)K} \cr\cr
      & \sim & \frac{k_B}{2}(1-2|r|)^2 K^2 e^{-(2|r|-1)K} 
\quad {\rm as} \ \ T \to 0 
\quad {\rm for} \ \ r < -\frac{1}{2} 
\label{crltmhalf}
\eeqs
while for the borderline value $r=-1/2$ one finds the proportionality 
\beq
C \sim k_B A K^2 e^{-2K} \quad {\rm for} \quad r = -\frac{1}{2}
\label{creqmhalf}
\eeq
For $r \ge 0$ and for physical temperature, the spin-spin correlation 
function decays asymptotically like
\beq
\langle \sigma_0 \sigma_n \rangle \sim \Bigl ( \frac{\lambda_{2+}}{
\lambda_{1+}} \Bigr )^{|n|}
\label{ss}
\eeq
This is also true for $-1/2 < r < 0$ if $T < T_D$ \cite{steph70,steph2}. 
Hence, 
taking the $T \to 0$ limit,  one finds that the correlation length $\xi$,
defined as usual by 
$\xi^{-1} = -\lim_{n \to \infty} n^{-1} \ln |\langle \sigma_0 \sigma_n 
\rangle|$, diverges like
\beq
\xi \sim (1/2) e^{2(1+2r)K} \quad {\rm as} \quad T \to 0
\label{xi}
\eeq
One also finds that for $r > -1/2$, the (zero-field) susceptibility $\chi$
diverges like 
\beq
\chi \sim (1/2)\beta^{-1}e^{2(1+2r)K} \quad {\rm as} \quad T \to 0
\label{chi}
\eeq
Consequently, for $r > -1/2$, 
\beq
C \sim K^2 \xi^{-1} \ , \quad \chi \sim K^{-1}\xi \quad {\rm as} 
\quad T \to 0 \ , 
\label{cxi}
\eeq
independent of $r$ in this range.  Thus, for $r > -1/2$, the singularities in
$C$ and $\chi$, expressed as functions of the correlation length $\xi$, are
independent of $r$ in this range, which shows that for $r > -1/2$, the NNN 
coupling is an irrelevant perturbation, and the model satisfies weak 
universality in the sense of Suzuki \cite{suzuki}, at the $T=0$ critical 
point.  

    In order to investigate which eigenvalues are dominant in various
complex-temperature phases, it is useful to express these as functions of the
Boltzmann weight variables.  For the case $r=1/p$ with integral $p$, since $Z$
is a generalized polynomial in $u$, the CT phase diagram is well defined in the
complex $u$ plane.  This follows since the CT zeros of $Z$ may be unambiguously
calculated in the $u$ plane, and, in the thermodynamic limit, these merge to
form the phase boundary ${\cal B}$. In terms of the variable $u$, 
\beq
\lambda_{1 \pm} = \frac{1}{2}\Bigl [ \ 1+u^{\frac{p}{2}} \pm 
\sqrt{(1-u^{\frac{p}{2}})^2 +4u^{1 + \frac{p}{2}}} \ \ \Bigr ] 
\label{lam1pmu}
\eeq
\beq
\lambda_{2 \pm} = \frac{1}{2}\Bigl [ \ 1-u^{\frac{p}{2}} \pm 
\sqrt{(1+u^{\frac{p}{2}})^2 -4u^{1 + \frac{p}{2}}} \ \ \Bigr ] 
\label{lam2pmu}
\eeq
Note that 
\beq
\lambda_{1\pm} \to \lambda_{2\pm} \quad {\rm for} \quad \sqrt{u} \to
-\sqrt{u}
\label{lamsym}
\eeq
If $p$ is an odd integer, then (\ref{lamsym}) implies that
\beq
\lambda_{1\pm}(u)=\lambda_{2\pm}(u^*) = \lambda_{2\pm}(u)^* 
 \quad {\rm for \ negative \ real} \ \ u
\label{uneg}
\eeq
respectively for the $\pm$ cases.  (Here, we use the standard branch cut for 
$\sqrt{u}$, along the negative real $u$ axis.)  

   As background, we recall that for the case of nearest-neighbor couplings,
${\cal B}$ consists 
of the negative real axis in the $u$ plane (e.g., Ref. \cite{is1d}).  This
is evident from the fact that for $r=0$, the two nontrivial eigenvalues of 
$\bar {\cal T}$
are $\lambda_{1+}=1+\sqrt{u}$ and $\lambda_{2+}=1-\sqrt{u}$, which are equal in
magnitude for negative real $u$.  (The other two eigenvalues, $\lambda_{1-}$
and $\lambda_{2-}$ both vanish.) 

\section{Case of $r=1/p$ for Positive Integer $p$}

   We first consider the situation where the NNN coupling is of the same sign
as, but weaker than, the NN coupling, i.e. $0 < r < 1$.  For our purposes, it 
will suffice to deal with the case where $r=1/p$
with $p$ a positive integer.  There are two subcases: $p$ even and $p$ odd. 
For even  $p=2\ell$, the eigenvalues of the transfer matrix 
have the following Taylor series expansions about $u=0$:
\beq
\lambda_{1+} = 1 + u^{1+\frac{p}{2}} + ...
\label{lam1ptayu}
\eeq
\beq
\lambda_{2+} = 1 - u^{1+\frac{p}{2}} + ...
\label{lam2ptayu}
\eeq
\beq
\lambda_{1-} = u^{\frac{p}{2}} + ...
\label{lam1mtayu}
\eeq
\beq
\lambda_{2-} = -u^{\frac{p}{2}} + ...
\label{lam2mtayu}
\eeq
where $...$ denote higher-order terms in $u$.
It follows that for even $p$, $\lambda_{1+}$ is the dominant eigenvalue on
the positive real $u$ axis and hence also in the complex-temperature phase
which includes this axis. Furthermore, on the negative real $u$ axis in the
vicinity of the origin, (i) if $p=0 \ {\rm mod} \ 4$, i.e., $\ell$ is even,
then $\lambda_{2+}$ is the dominant eigenvalue, whereas
(ii) if $p = 2 \ {\rm mod} \ 4$, i.e., $\ell$ is odd, then $\lambda_{1+}$ is
the dominant eigenvalue; in both cases, the respective eigenvalues are
therefore also dominant in the CT phases which include this portion of the
negative real $u$ axis near the origin. 
For odd positive integral $p$, the $\lambda$'s have analogous series
expansions in the $z$ plane, $\lambda_{1+} = 1 + z^{2+p} + ...$, etc. 
Hence, in the $u$ plane (with the usual $+$ sign taken for $\sqrt{u}$ if
$\arg(u)=0$), $\lambda_{1+}$ is again dominant on the positive real $u$ axis
in the vicinity of the origin. 

   Together with the theorem that a 1D spin model
with finite-range interactions has no non-analyticity for $T > 0$, i.e.,
along the positive $u$ axis, it follows that for positive integral $p$, 
$\lambda_{1+}$ is the dominant eigenvalue on the entire positive $u$ axis and 
hence the CT phase in the $u$ plane which includes this axis and to which one
can thus analytically continue from this axis.  

    We next prove a general theorem: For $r=1/p$ with $p$ a positive integer, 
there are $p+2$ phase boundary curves emanating from the origin in the 
complex $u$ plane, at the angles 
\beq
\theta_n = \frac{(2n+1)\pi}{2+p} \ , \qquad {\rm for} \quad n=0,...,p+1
\label{thetan}
\eeq
Proof: To encompass the cases of both even and odd $p$, we use the Taylor 
series expansions in the $z$ plane.  From these, it follows that in the 
vicinity of $z=0$, $\lambda_{1+}$ and $\lambda_{2+}$ alternate as the dominant 
eigenvalues.  Now define polar coordinates according to 
$z=\rho_z e^{i\theta_z}$, whence $u=\rho e^{i\theta}$ with $\rho=\rho_z^2$ and
$\theta=2\theta_z$; then the degeneracy condition of leading eigenvalues, 
$|\lambda_{1+}| = |\lambda_{2+}|$, reads $|1+z^{2+p}+...| = |1-z^{2+p}+...|$ 
(where $...$ denote higher-order terms), the solution to which is 
$\cos\Bigl ( (2+p)\theta_z \Bigr ) =0$, i.e.,
$\theta_z=(2n+1)(\pi/2)/(2+p)$ for $n=0,..., p+1$.  This proves
 eq. (\ref{thetan}). 

   A related theorem is: For positive, odd $p$, the complex-temperature phase
boundary ${\cal B}$ always contains the negative real $u$ axis.  To
prove this, we again use the result that in the vicinity of $z=0$, 
$\lambda_{1+}$ and $\lambda_{2+}$ alternate as the dominant eigenvalues.  We
next observe that in the $u$ plane, the degeneracy condition of leading 
eigenvalues, $|\lambda_{1+}| = |\lambda_{2+}|$ is automatically satisfied on 
the negative real $u$ axis as a consequence of the symmetry condition 
(\ref{lamsym}) and relation (\ref{uneg}). This completes the proof. 

   Two further general theorems are the following: For positive $r=1/p$ with 
integer $p$, as one
makes a half-circuit around the origin in the $u$ plane, the dominant 
eigenvalues alternate between $\lambda_{1+}$ and $\lambda_{2+}$.  This is 
proved by noting first that from the Taylor series expansions above, these 
two eigenvalues are the dominant ones in the vicinity of the origin, and 
second, it is precisely their alternation as dominant eigenvalues 
which produces the phase boundaries emanating from the origin at the angles 
(\ref{thetan}) and separating the different phases.  
Since the dominant eigenvalue at $-\theta$ is the same as
that at $\theta$, this theorem also completely determines the dominant
eigenvalues on the rest of the full circle around the origin. 
 
\begin{figure}
\epsfxsize=3.5in
\epsffile{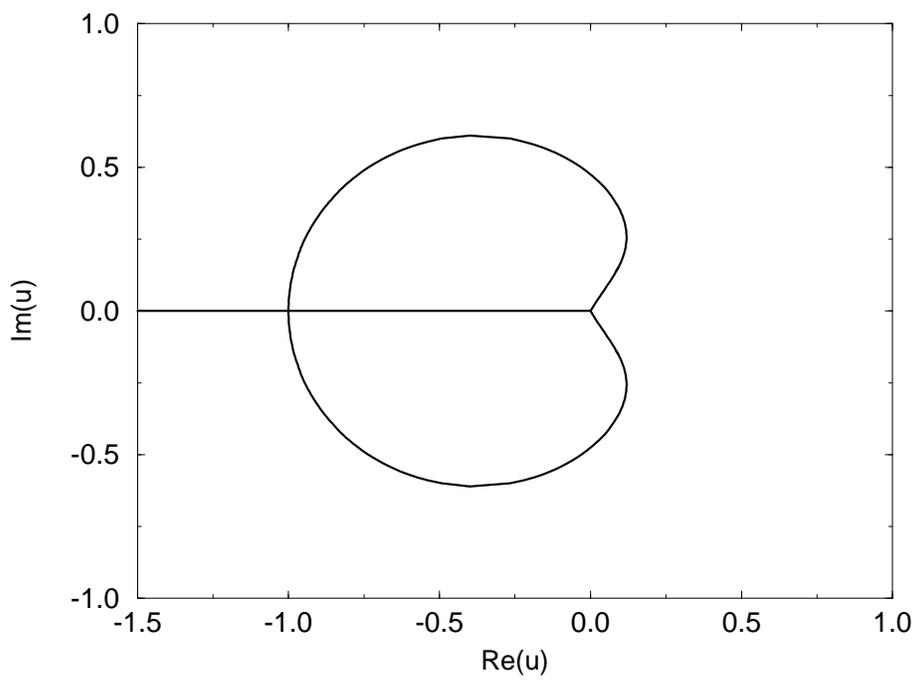}
\caption{Phase diagram of the 1D NNN Ising model in the complex $u$
plane for $J_{nnn}/J_{nn} = r =1$.}
\label{figr1}
\end{figure}

\begin{figure}
\epsfxsize=3.5in
\epsffile{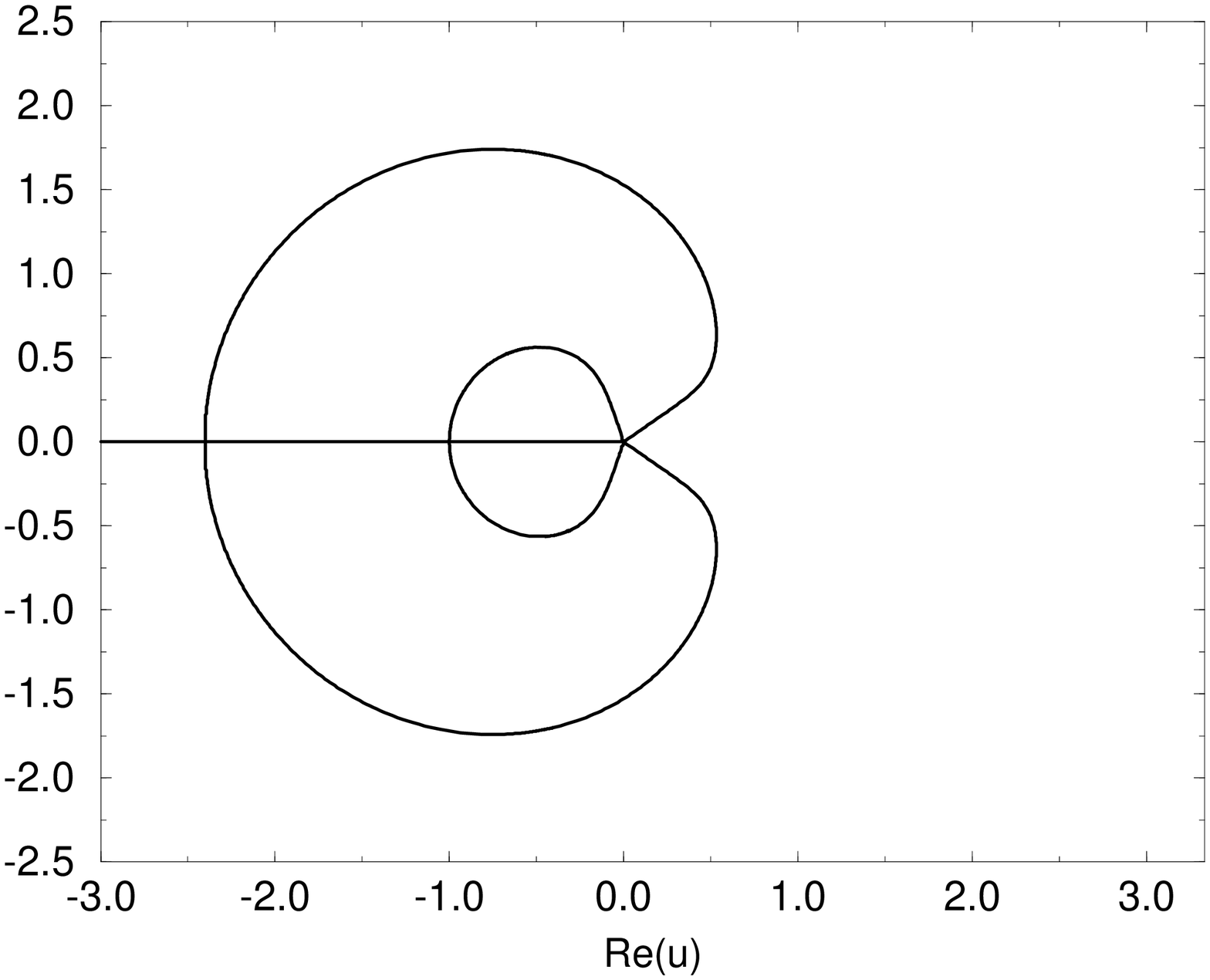}
\caption{As in Fig. \ref{figr1}, for $r=1/3$.} 
\label{figr1ov3}
\end{figure}

\begin{figure}
\epsfxsize=3.5in
\epsffile{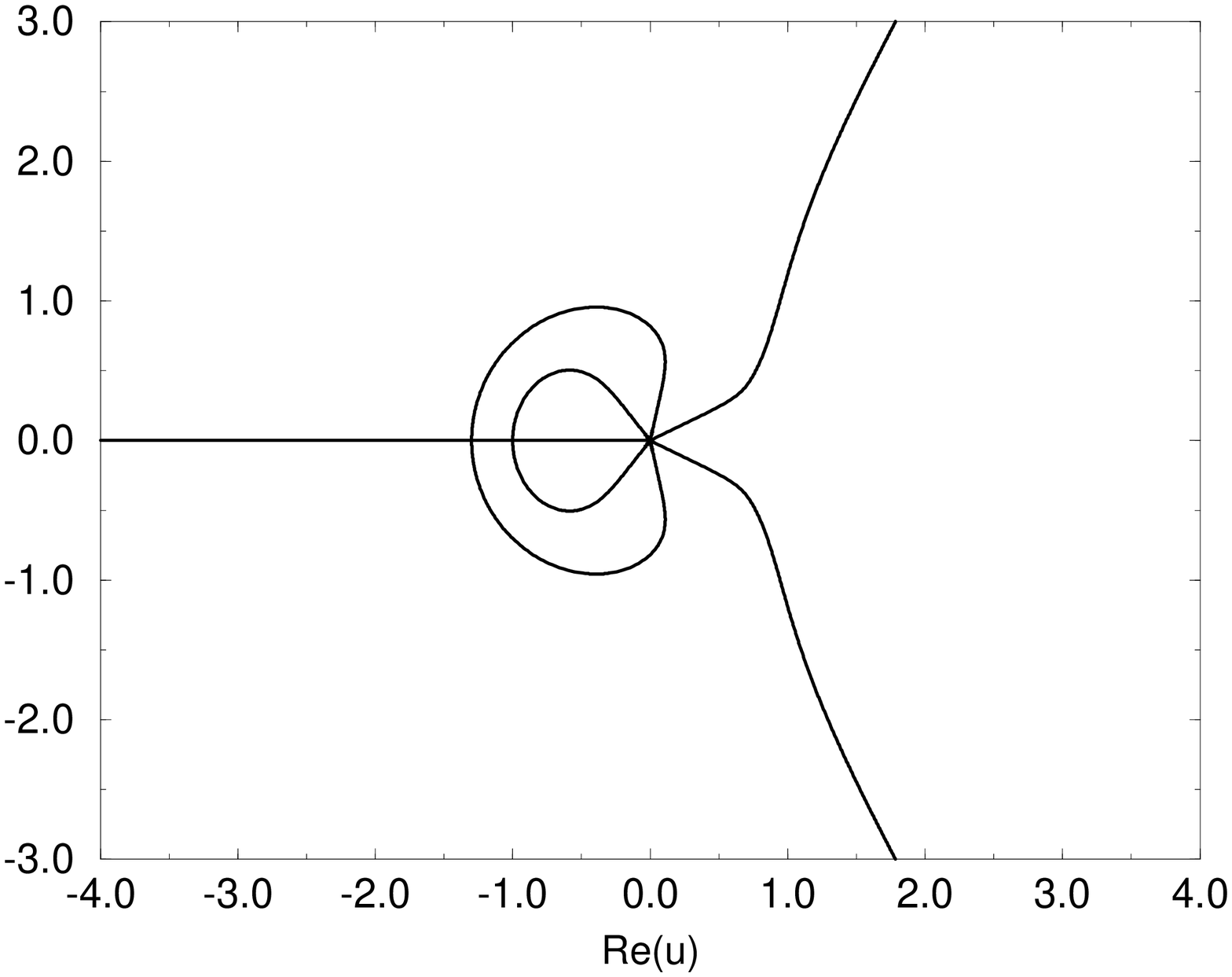}
\caption{As in Fig. \ref{figr1}, for $r=1/5$.}
\label{figr1ov5}
\end{figure}

   By solving the degeneracy conditions of dominant eigenvalues, we have mapped
out the complex-temperature phase diagrams.  We consider odd values of $p$
first and then even values. In Figs. \ref{figr1}-\ref{figr1ov5} we show the 
results for $r=1$, $1/3$, and $1/5$. 

For $r=1$, we find that the complex-temperature 
phase diagram consists of three phases: (a) a region
including the positive real $u$ axis and extending outward to the circle at
infinity, together with two complex-conjugate (c.c) phases (b), (b)$^*$,
located above and below the negative real $u$ axis from the origin leftward to
$u=-1$.  As follows from our general discussion above, $\lambda_{1+}$ is the
dominant eigenvalue in region (a), and, since $\lambda_{1+}$ and $\lambda_{2+}$
alternate as dominant eigenvalues in the vicinity of $u=0$, $\lambda_{2+}$ is
dominant in regions (b) and (b)$^*$. On 
curve separating region (a) from regions (b) and (b)$^*$,
$|\lambda_{1+}|=|\lambda_{2+}|$.  The CT phase boundary ${\cal B}$ includes 
multiple points at $u=0$, where three curves meet, and a multiple point at
$u=-1$, where four curves meet, with 2 different tangents, hence index 2. 
Here, we use the term ``multiple point'' and ``index'' in their technical 
algebraic geometry sense (see our previous discussions in Ref. \cite{cmo}). 
Thus, while in the model with only the NN spin-spin coupling, the CT phase 
boundary ${\cal B}$ is the negative real axis, the effect of adding a 
NNN spin-spin coupling with $r=1$ is to produce two new 
complex-conjugate phases
bounded by curves starting out from the origin and meeting at
$u=-1$.  As must be true from general arguments, the theory is still analytic
on the positive real $u$ axis; the only changes are an increase in the number
of CT phases elsewhere.

  The complex-temperature phase diagram for $r=1/3$, Fig. \ref{figr1ov3}, 
is progressively more
complicated, consisting of five phases: (a) a region containing the positive 
real $u$ axis; (b), (b)$^*$, complex-conjugate phases whose borders are 
shaped somewhat like half-circles, adjacent to the negative $u$ axis and 
including the interval $-1 < u < 0$; and (c), (c)$^*$, c.c. phases lying
roughly concentrically outward from (b), (b)$^*$, and including the interval of
the negative real $u$ axis $-2.4 \ltwid u < -1$.  From our general discussion
above, it follows that, in addition to region (a), 
$\lambda_{1+}$ is dominant in regions
(b) and (b)$^*$ while $\lambda_{2+}$ is dominant in regions (c) and (c)$^*$. 
The CT phase boundary ${\cal B}$ contains multiple points at $u=0$ where 
five curves come together, at $u=-1$ and $u \simeq -2.4$, where in each case 
four curves come together with two different tangents, hence index 2. 

   Finally, we show the complex-temperature phase diagram for $r=1/5$ in
Fig. \ref{figr1ov5}.  The general features of this phase diagram follow
 from our previous discussion.  A new aspect is that in addition to the part of
${\cal B}$ running along the negative real $u$ axis, ${\cal B}$ also contains
two complex-conjugate curves which extend to infinite distance from the origin
in the ``northeast'' and ``southeast'' quadrants. 

\begin{figure}
\epsfxsize=3.5in
\epsffile{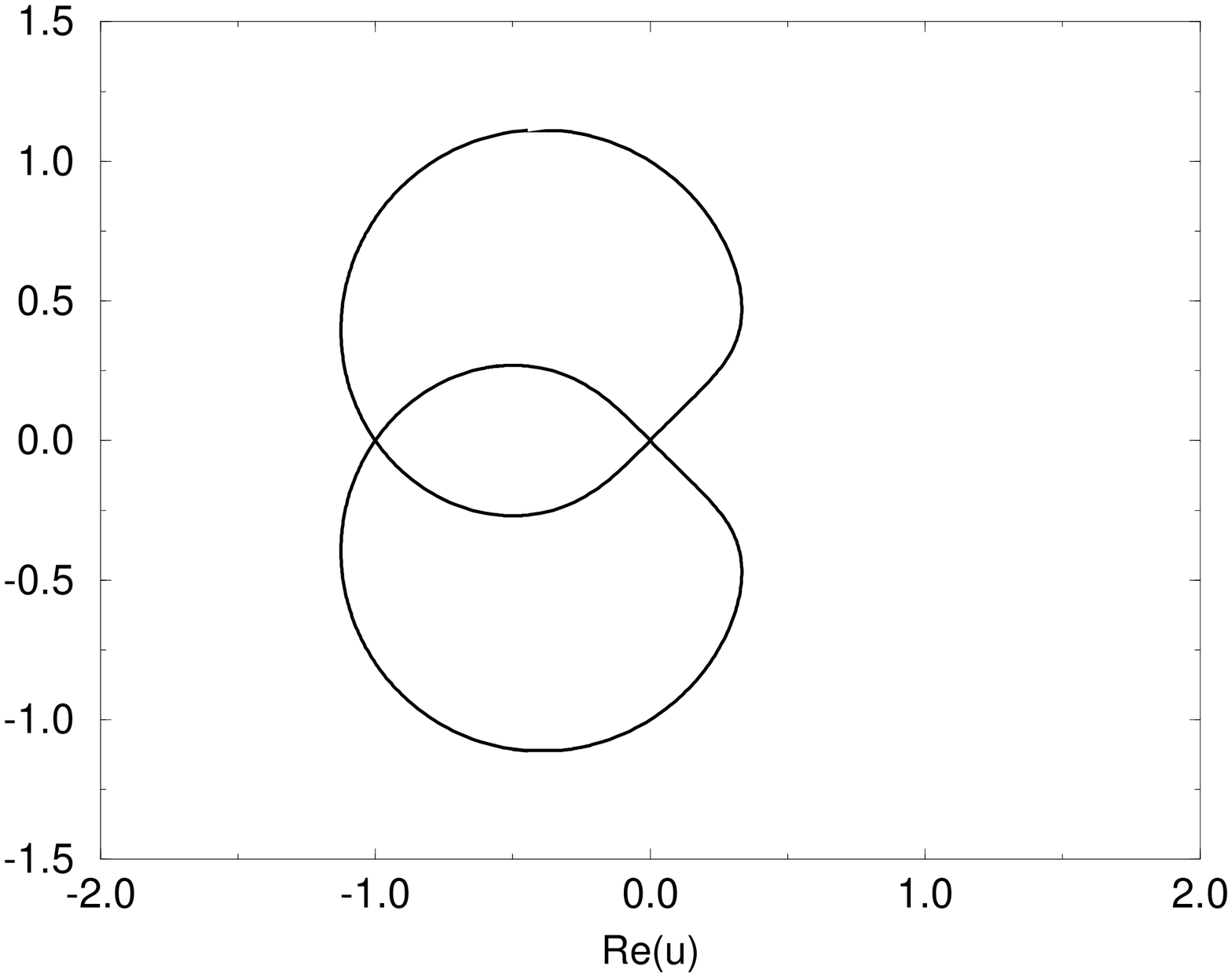}
\caption{As in Fig. \ref{figr1}, for $r=1/2$.}
\label{figr1ov2}
\end{figure}

   We next show in Fig. \ref{figr1ov2} a typical CT 
phase diagram in the $u$ plane for an even value, $p=2$, i.e., $r=1/2$. 
In contrast to the diagrams with odd $p$, in those with even $p$, 
${\cal B}$ does not contain the negative real axis.  
For $p=2$, the CT phase diagram consists of four phases: (a) a region 
containing the positive real $u$ axis and extending outwards to infinity; 
(b) a region including the interval $-1 < u < 0$; and complex conjugate 
regions (c), (c)$^*$ above and below region (b).  Our general discussion above
determines the dominant eigenvalues in the various regions: $\lambda_{1+}$ in
(a) and (b), and $\lambda_{2+}$ in regions (c) and (c)$^*$. The CT phase 
boundary ${\cal B}$ involves multiple points at $u=0$ and $u=-1$, each of 
index 2.  The phase boundaries for even $p$ may run to infinite distance from
the origin.  For example, we have also calculated the phase diagram for 
$p=4$ case and find in this case that part of ${\cal B}$ consists of curves
running to $u = \pm i \infty$. 

\section{Case of Positive Integer $r$} 

   We have also studied the situation where the NNN coupling is ferromagnetic
and stronger than the NN coupling, i.e. $r \ge 1$; here, we focus on the case 
of positive integer $r$.
For this case, the partition function is a generalized polynomial in $u_{_K}$. 
By re-expressing the eigenvalues $\lambda_{1\pm}$ and
$\lambda_{2\pm}$ as functions of $u_{_K}$, i.e., 
\beq
\lambda_{1 \pm} = \frac{1}{2}\Bigl [ \ 1+u_{_K}^{\frac{1}{2}} \pm
\sqrt{(1-u_{_K}^{\frac{1}{2}})^2 +4u_{_K}^{\frac{1}{2}+r}} \ \ \Bigr ]
\label{lam1pmuk}
\eeq
\beq
\lambda_{2 \pm} = \frac{1}{2}\Bigl [ \ 1-u_{_K}^{\frac{1}{2}} \pm
\sqrt{(1+u_{_K}^{\frac{1}{2}})^2 -4u_{_K}^{\frac{1}{2}+r}} \ \ \Bigr ]
\label{lam2pmuk}
\eeq
one sees that 
\beq
|\lambda_{1+}(u_{_K})|=|\lambda_{2+}(u_{_K})| \ , \quad 
|\lambda_{1-}(u_{_K})|=|\lambda_{2-}(u_{_K})|
\quad {\rm for \ \ real} \quad 
u_{_K} < 0
\label{lamreluk}
\eeq
This implies that the CT phase boundary ${\cal B}$ contains the negative real
$u_{_K}$. The expansions of these four eigenvalues around the
origin of the $u_{_K}$ plane follow directly from the expansions given in
eqs. (\ref{lam1ptayu})-(\ref{lam2mtayu}) with the replacement $u = u_{_K}^r$.
Hence, it is again true that as one traverses a half-circuit of the origin in
the $u_{_K}$ plane, the dominant eigenvalue along the positive real 
$u_{_K}$ axis is
$\lambda_{1+}$ and the dominant eigenvalues alternate between $\lambda_{1+}$
and $\lambda_{2+}$.  This also determines the dominant eigenvalues on the
complex-conjugate half-circuit.  These two results together imply that for
arbitrary positive integral $p$, the CT phase boundary ${\cal B}$ always 
includes the negative real $u_{_K}$ axis.  

   Re-expressing the Taylor series expansions of the eigenvalues in terms of 
$z_{_K}$, the degeneracy condition for the leading eigenvalues, 
$|\lambda_{1+}|=|\lambda_{2+}|$ reads 
$|1+z_K^{1+2r}+ ...| = |1-z_K^{1+2r}+...|$, where $...$ denote higher-order
terms.  Denoting $u_{_K} = \rho_{_K}e^{i\theta_{_K}}$, the solution to this
condition is 
\beq
\theta_{_K} = \frac{(2n+1)\pi}{1+2r}  \quad {\rm for} \quad 
n = 0,...2r
\label{thetauk}
\eeq
This proves that in the complex-temperature phase diagram in the $u_{_K}$ 
plane, ${\cal B}$ contains $1+2r$ curves emanating from
the origin at the angles given in eq. (\ref{thetauk}). 

\begin{figure}
\epsfxsize=3.5in
\epsffile{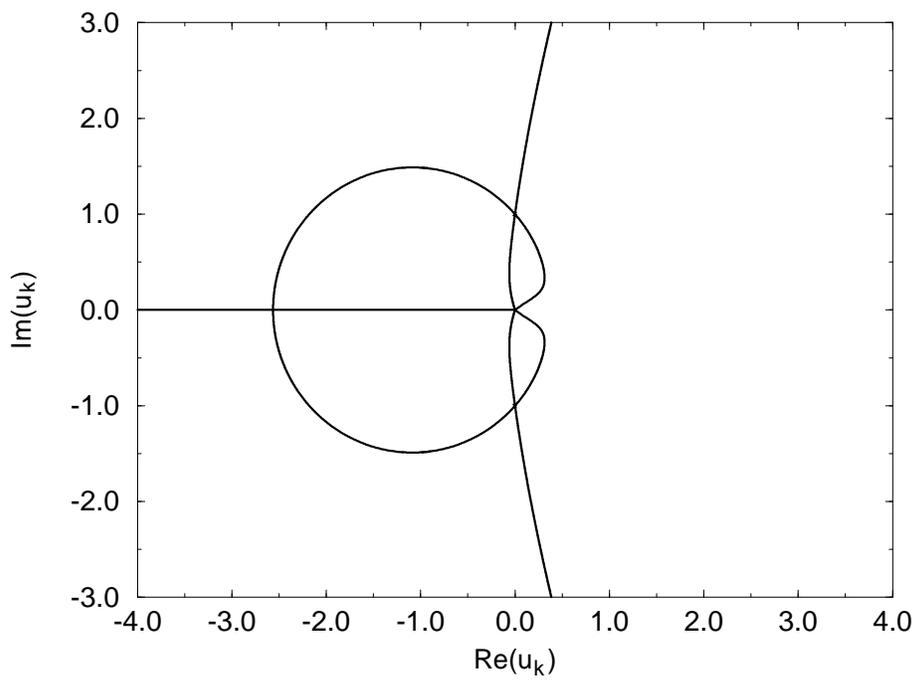}
\caption{Complex-temperature phase diagram in the $u_{_K}$ plane for $r=2$.}
\label{figr2}
\end{figure}

   In Fig. \ref{figr2} we show the complex-temperature phase diagram for a
typical case, $r=2$. 
The dominant eigenvalues in the phases which are contiguous to the origin 
$u_{_K}=0$ are completely determined by our previous general results; 
starting from the phase containing the positive real $u_{_K}$ axis and moving 
in the direction of increasing $\arg(u_{_K})$, these
alternate according to $\lambda_{1+}$, $\lambda_{2+}$, and $\lambda_{1+}$. For 
the remaining phase which is not contiguous to the origin, depending on one's
choices of branch cuts connecting the branch points of the square roots in the
eigenvalues, either $\lambda_{2+}$ or $\lambda_{2-}$ is dominant.

\section{Negative $r$ in the Interval $-1/2 < r < 0$}

   For the range $-1/2 < r < 0$, notwithstanding the competition and 
frustration which the NNN interaction produces, it is still an irrelevant 
perturbation.  As before, if $1/r=-p$ with positive integral $p$, then 
$Z$ is a generalized polynomial in the variable $u=e^{-4K'}$ and hence also in
its inverse, $w=1/u$. Since $K' < 0$, we shall use this inverse variable $w$ 
for our analysis, in order to maintain the correspondence of zero temperature 
with the origin of the plots.  The eigenvalues of the
transfer matrix $\bar {\cal T}$ in this variable can be obtained from
eqs. (\ref{lam1pmu}) and (\ref{lam2pmu}):
\beq
\lambda_{1 \pm} = \frac{1}{2}\Bigl [ 1 + w^{\frac{p}{2}} 
\pm \sqrt{ (1-w^{\frac{p}{2}})^2 + 4w^{\frac{p}{2}-1} } \ \ \Bigr ]
\label{lam2pw}
\eeq
\beq
\lambda_{2 \pm} = \frac{1}{2}\Bigl [ 1 - w^{\frac{p}{2}} 
\pm \sqrt{ (1+w^{\frac{p}{2}})^2 - 4w^{\frac{p}{2}-1} } \ \ \Bigr ]
\label{lam2pmw}
\eeq
The borderline value $p=2$, i.e., $r=-1/2$ at which the nature of the ground
state changes, as discussed above, is evident in these eigenvalues since as $p$
increases above $p=2$, i.e., $r$ decreases below $r=-1/2$, the eigenvalues
cease to be finite at the origin $w=0$ because the last term in the square root
becomes a negative power.  By the same reasoning as before, if and only if 
$p$ is an odd integer, ${\cal B}$ contains the negative real $w$ axis. 
For integer $p > 2$, so that $r > -1/2$, which 
includes the region of interest here, these eigenvalues have the 
series expansions around $w=0$ 
\beq
\lambda_{1+} = 1 + w^{\frac{p}{2}-1} + ...
\label{lam1pwtay}
\eeq
\beq
\lambda_{2+} = 1 - w^{\frac{p}{2}-1} + ...
\label{lam2pwtay}
\eeq
\beq
\lambda_{1-} = -w^{\frac{p}{2}-1} + ...
\label{lam1mwtay}
\eeq
\beq
\lambda_{1+} = w^{\frac{p}{2}-1} + ...
\label{lam2mwtay}
\eeq
where $...$ denote higher order terms. 
 From this it follows that for our case of integer $p > 2$, $\lambda_{1+}$ is
the dominant eigenvalue on the positive real $w$ axis and in the CT phase which
includes this axis.  Other results are similar to those derived for positive
$r=1/p$ above: on the negative real $w$ axis in the vicinity of the origin, 
if $p$ is even, then (i) if $p=0 \ {\rm mod} \ 4$, then $\lambda_{2+}$ is the
dominant eigenvalue, whereas (ii) if $p = 2 \ {\rm mod} \ 4$, then
$\lambda_{1+}$ is dominant, and these respective eigenvalues are also dominant
in the CT phase which includes this portion of the negative real $w$ axis. 

\begin{figure}
\epsfxsize=3.5in
\epsffile{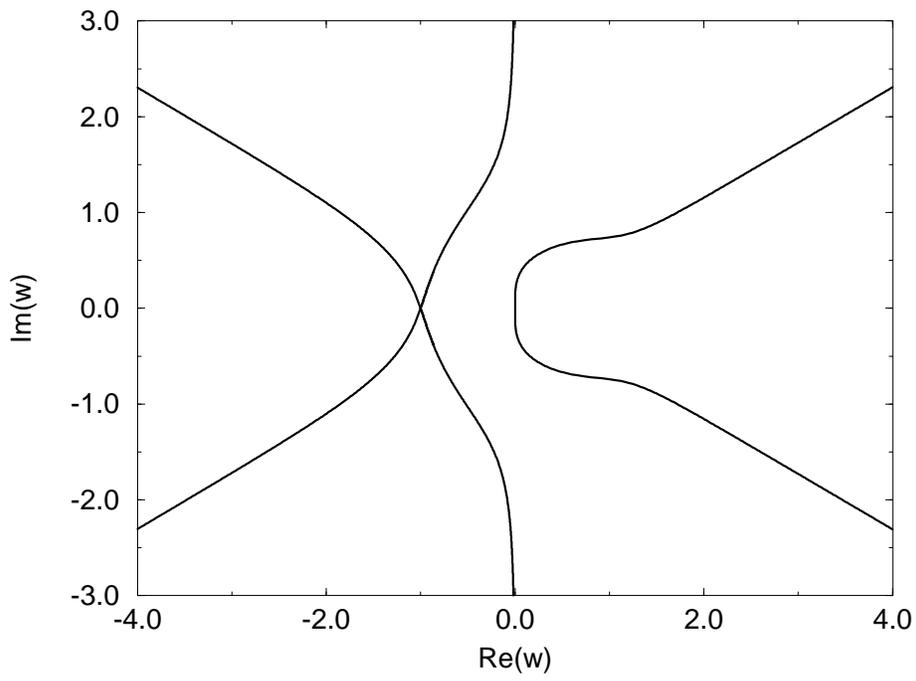}
\caption{Complex-temperature phase diagram in the $w$ plane for $r=-1/4$.}
\label{figrm1ov4}
\end{figure}

\begin{figure}
\epsfxsize=3.5in
\epsffile{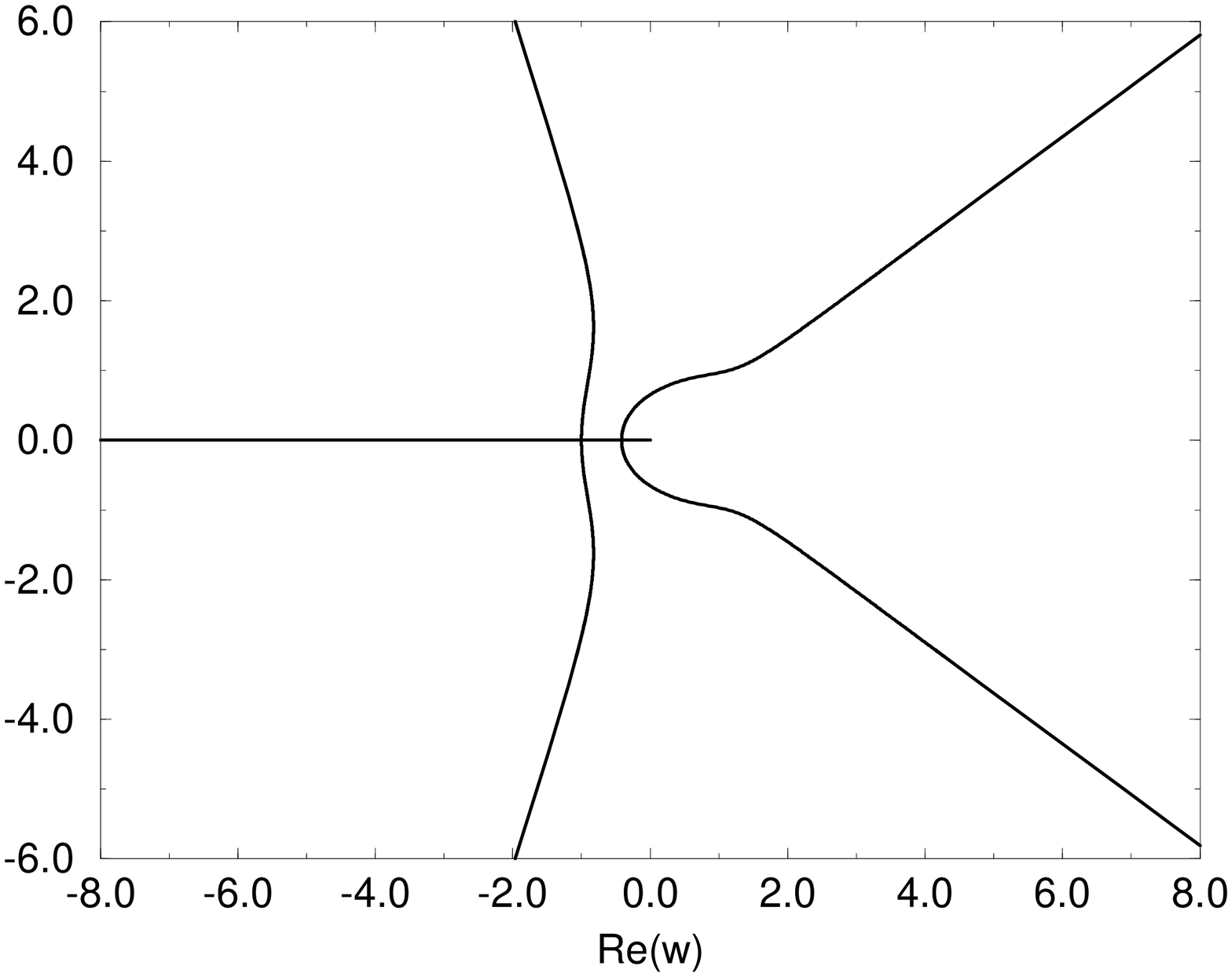}
\caption{Complex-temperature phase diagram in the $w$ plane for $r=-1/3$.}
\label{figrm1ov3}
\end{figure}

In Fig. \ref{figrm1ov4} and \ref{figrm1ov3} we show our calculation of the 
complex-temperature phase diagram in the $w$ plane for the values $r=-1/4$ and
$r=-1/3$, respectively. 
For $r=-1/4$, besides (a) the wedge-shaped phase including the positive real 
$w$ axis, where $\lambda_{1+}$ is dominant, there is a phase (b) which 
includes the interval $-1 < w < 0$ of the $w$ axis, in which $\lambda_{2+}$ 
dominates, and (c) a phase including the segment $-\infty < w < -1$ where
$\lambda_{2-}$ is dominant.  In addition, there are two complex-conjugate
phases (d) and (d)$^*$ where $\lambda_{1-}$ is dominant; 
these have boundaries which cross each
other at a multiple point of index 2 at $w=-1$. 
For $r=-1/3$, besides the phase containing the positive real $w$ axis, there
are two pairs of complex-conjugate phases.  As one moves from northeast to
northwest, the eigenvalues which are dominant in these regions are
$\lambda_{2+}$ and $\lambda_{1-}$. 

   For the present range $-1/2 < r < 0$, there is a finite physical disorder
temperature $T_D$ determined by the equation \cite{steph70,steph2} 
\beq
\cosh(K) = e^{-2K'}
\label{tdcurve}
\eeq
The disorder temperature $T_D$ decreases monotonically from 
$T_D=\infty$ at $r=0$ to $T_D=0$ as $r$ decreases to $-1/2$.  In the context of
the complex-temperature generalization of this model, we observe that, in
addition to the physical solution of (\ref{tdcurve}) for $T_D$, there are also
complex-temperature solutions.  In terms of the ratio $r$ and the coupling 
$K = K_R + iK_I$, the real and imaginary parts of eq. (\ref{tdcurve}) yield 
the respective equations
\beq
\cosh(K_R)\cos(K_I) = e^{-2rK_R}\cos(2rK_I)
\label{retdcurve}
\eeq
\beq
\sinh(K_R)\sin(K_I) = -e^{-2rK_R}\sin(2rK_I)
\label{imtdcurve}
\eeq
A more compact way of writing (\ref{tdcurve}) is in terms of the Boltzmann
variable $z_{_K}$:
\beq
1+z_{_K} = 2z_{_K}^{r+\frac{1}{2}}
\label{diseq}
\eeq
Let us define $K_D = J_{nn}/(k_BT_D)$ and $z_{_K}=e^{-2K_D}$. 
As an illustration, for $r=-1/4$, eq. (\ref{diseq}), expressed in terms of 
the variable $\omega = w^{1/2}$, where $w =e^{4K'}$ (whence, 
$w(r=-1/4) = z_{_K}^{1/2} = e^{-K}$) 
is $(\omega - 1)(\omega^3 + \omega^2 + \omega - 1)=0$. (The trivial solution 
$\omega=1$ corresponds to $K = K'=0$ in (\ref{tdcurve}) and is not of 
interest here.)  The cubic factor has as roots the physical disorder solution
$\omega_D = 0.5437$, i.e., $w_D=0.2956$ \ ($K_D = 1.219$) and, in
addition, the complex-temperature roots 
$\omega = -0.7718 \pm 1.115i$, i.e., $w = -0.6478 \pm 1.721i$ \ 
($K = -0.6094 \pm 1.931i$).  As noted in Ref. \cite{ms}, there are 
an infinite number of complex $K$ values corresponding to a given value of a 
Boltzmann weight variable, depending on one's choice of Riemann sheet in the
evaluation of the logarithm; here, we list only one value of $K$ for each 
$w$. These complex-temperature solutions of 
(\ref{tdcurve}) lie in the phases (d) and (d)$^*$ in 
Fig. \ref{figrm1ov4}.  From a similar analysis for $r=-1/3$, where $w(r=-1/3)
= e^{-4K/3}$, we find, besides the physical disorder point 
$w_D=0.06694$ \ ($K_D = 2.028$), also the two pairs of complex-temperature
solutions $w=-1.607 \pm 1.539i$ \ ($K=-0.5998 \pm 1.783i$) and 
$w=1.073 \pm 1.366i$ \ ($K=-0.4142 \pm 0.6787i$).  One can see from
Fig. \ref{figrm1ov3} that these complex-temperature solutions of 
(\ref{tdcurve}) lie in the interiors of four CT phases.

\section{Negative $r$ in the Interval $-\infty < r < -1/2$} 

\begin{figure}
\epsfxsize=3.5in
\epsffile{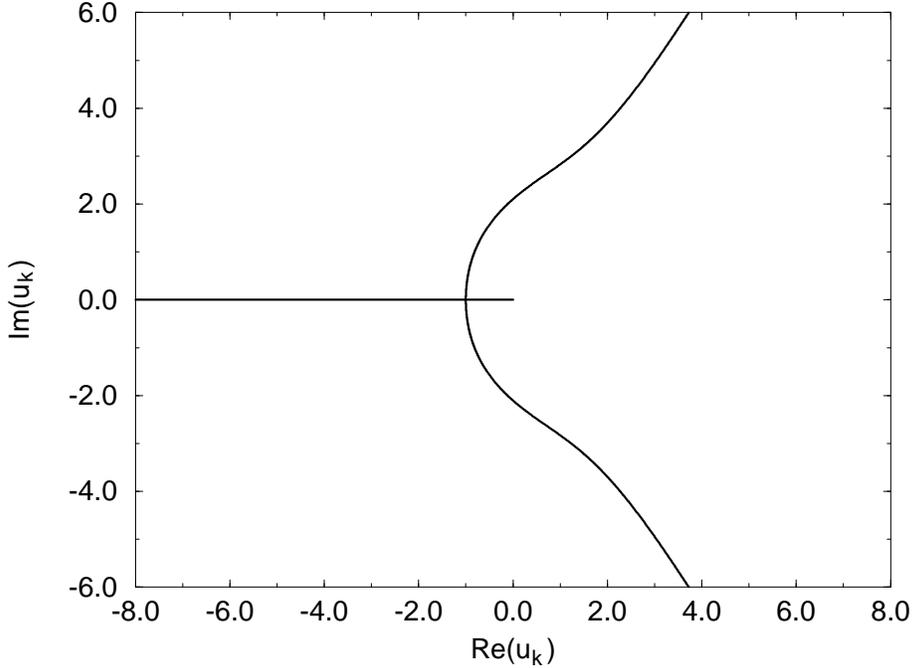}
\caption{Complex-temperature phase diagram in the $u_{_K}$ plane for $r=-1$.}
\label{figrm1}
\end{figure}

   As discussed above, for $r \le -1/2$, the NNN spin-spin coupling is so
strong as to change the nature of the ground state from FM to the $(2,2)$
form. In Fig. \ref{figrm1} we show the complex-temperature phase diagram for a 
typical case, $r=-1$.  Note that, in
particular, by the same reasoning as for positive integer $r$ (c.f. eq. 
(\ref{lamreluk})), it follows that ${\cal B}$ always includes the negative real
$u_{_K}$ axis. 

\section{Case $r=-1/2$}

\begin{figure}
\epsfxsize=3.5in
\epsffile{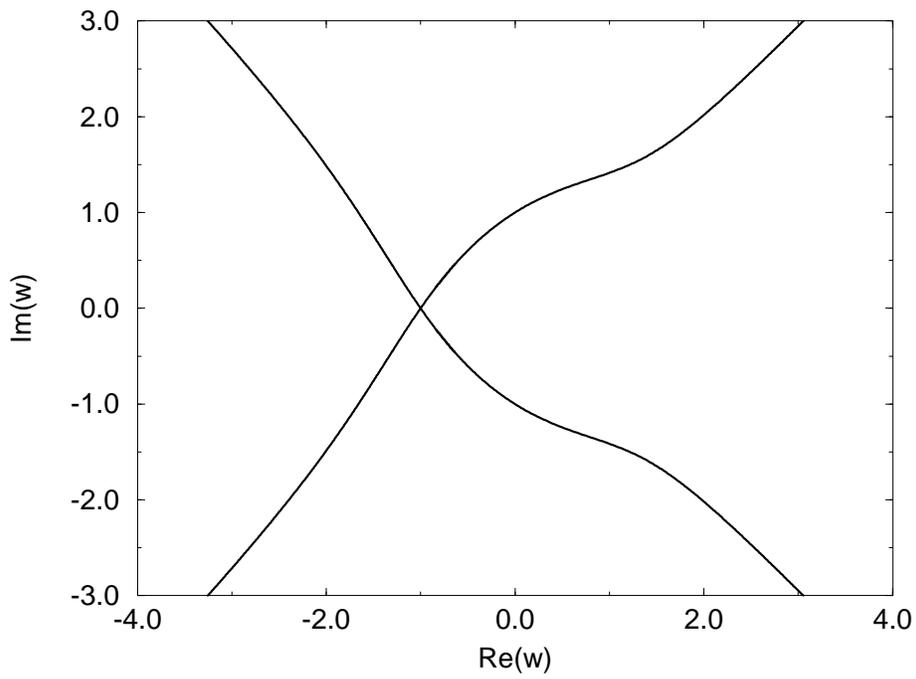}
\caption{Complex-temperature phase diagram in the $w$ plane, for $r=-1/2$.}
\label{figrm1ov2}
\end{figure}

   For the borderline value $r=-1/2$, i.e., $J_{nnn}=-(1/2)J_{nn} < 0$, 
the competing preferences toward a ferromagnetic and (2,2) ground state are 
exactly balanced.  Indeed, for $r=-1/2$, the model has nonzero ground 
state entropy, $S(T=0) = k_B \ln \Bigl \{ (1/2)(1+\sqrt{5}) \Bigr \}$
\cite{steph70,redner}, and exponential asymptotic decay of $\langle \sigma_0
\sigma_n \rangle$ (modulated by an oscillatory factor) even at $T=0$
\cite{steph70,steph2}.  We find that if and only if 
$r=-1/2$, then the complex-temperature phase boundary ${\cal B}$ does not pass 
through the point $T=0$, or equivalently, the origin in the $w=1/u=e^{4K'}$ 
plane.  This avoidance of the point $w=0$ by ${\cal B}$ shows the absence of 
criticality at $T=0$.  In Fig. \ref{figrm1ov2} we present our calculation of 
the complex-temperature phase diagram for $r=-1/2$.  One sees that ${\cal B}$
consists of two complex-conjugate curves which only intersect the real $w$ axis
at the point $w=-1$ (where they exhibit a multiple point of index 2). 

\section{Potts Model} 

      Since the spin 1/2 Ising model is equivalent to the two-state Potts 
model, it is natural to extend the present study to include some remarks on how
the complex-temperature phase diagram of the 1D Potts model changes under the
addition of a NNN coupling which is an irrelevant operator.  Recall that in
contrast to the 2D NN (spin 1/2) Ising model, no exact solution is known for 
general temperature of the 2D NN $q$ state Potts model for $q > 2$ and hence 
the CT boundary ${\cal B}$ is not known even
for this NN case (see, e.g., Refs. \cite{chw,pfef} and references therein).

The zero-field 1D $q$-state Potts model with NN and NNN interactions 
is defined by the partition function 
$Z_P = \sum_{\sigma_n}e^{-\beta {\cal H}_P}$ with
\beq
{\cal H}_P = -J_{nn} \sum_n\delta_{\sigma_n \ \sigma_{n+1}} - 
J_{nnn} \sum_n\delta_{\sigma_n \ \sigma_{n+2}} 
\label{hpotts}
\eeq
where $\delta_{ij}$ is the Kronecker delta and $\sigma_n \in \{1,...,q\}$.  
For our study, we shall consider 
ferromagnetic couplings, $J_{nn}, \ J_{nnn} > 0$ and define $K$, $K'$,
$r$ as in eqs. (\ref{k})-(\ref{r}), and $u_{_P}=e^{-K}$. 
The 1D NN model with $J_{nn} < 0$ involves finite ground state 
disorder, with ground state entropy $S_0= k_B \ln(q-1)$. This is similar to the
situation on several higher-dimensional lattices; see Ref. \cite{baxter} for
the square lattice and Ref. \cite{st} for the honeycomb lattice. 

   A notable feature of the CT phase diagram for the 1D NN Potts model is its 
simplicity; in Ref. \cite{is1d} we found that for 
$q \ge 3$, it consists only of two phases, separated by the boundary 
${\cal B}$ comprised of a circle 
\beq
u_{_P} = \frac{-1 + e^{i\omega}}{q-2} \ , \quad 0 \le \omega < 2\pi
\label{ucircle}
\eeq

   The solution of the 1D NNN Potts model proceeds in the standard manner, 
via transfer matrix methods.  As before, it is most efficient to use spin
configuration vectors $v_n=|\sigma_n, \sigma_{n+1}\rangle$, so that 
\beq
\langle v_n|{\cal T_P}|v_{n+1} \rangle =
\langle v_n |e^{-\beta {\cal H}}|v_{n+1}\rangle =
\exp \Bigl ( \frac{K}{2}(\delta_{\sigma_n \ \sigma_{n+1}} + 
\delta_{\sigma_{n+1} \ \sigma_{n+2}} )
+ K'\delta_{\sigma_n \ \sigma_{n+2}} \Bigr )
\label{tpotts}
\eeq
Thus the transfer matrix ${\cal T}_P$ is a $q^2 \times q^2$ matrix.  Here we 
consider the simplest case, $q=3$.  The resultant ${\cal T}_P$ is 
straightforwardly calculated from (\ref{tpotts}).  
Defining $\bar {\cal T}_P = e^{-(K+K'+h)}{\cal T}_P$, we find for the 
characteristic polynomial of $\bar {\cal T}_P$: 
\begin{eqnarray}
P(\bar {\cal T}_P;\lambda) & = & \Bigl ( \lambda+u_{_P}(1-u_{_P}) \Bigr ) 
\Bigl ( \lambda^2 - (u_{_P}^2+u_{_P}+1)\lambda + u_{_P}(1-u_{_P})(1+2u_{_P}) 
\Bigr ) \times
\cr 
& & \times \Bigl ( \lambda^3 + (u_{_P}^2-1)\lambda^2
+u_{_P}^2(u_{_P}-1)(u_{_P}+2)\lambda 
+ u_{_P}^2(1-u_{_P})^2(2u_{_P}+1) \Bigr )^2 
\label{charpolpotts}
\end{eqnarray} 
The resultant eigenvalues of $\bar {\cal T}_P$ are
\beq
\lambda_0 = u_{_P}(u_{_P}-1)
\label{lam1potts}
\eeq
\beq
\lambda_{1 \pm} = \frac{1}{2}\Bigl [ 1 + u_{_P} + u_{_P}^2 \pm 
\sqrt{ 1-2u_{_P}-u_{_P}^2+10u_{_P}^3+u_{_P}^4} \ \ \Bigr ]
\label{lam2pmpotts}
\eeq
together with three roots of the cubic factor in (\ref{charpolpotts}), each of
which is a double root of $P(\bar {\cal T}_P;\lambda)$.  We denote these as 
$\lambda_{3a}$ $\lambda_{3b}$, and $\lambda_{3c}$ and, since the expressions 
for these cubic roots are rather complicated, we omit listing them here.  
Aside from the polynomial $\lambda_0$, the other eigenvalues have the following
Taylor series expansions around $u_{_P}=0$:
\beq
\lambda_{1+} = 1 + 2u_{_P}^3 + O(u_{_P}^4) 
\label{lam1ptay}
\eeq
\beq
\lambda_{1-} = u_{_P} + u_{_P}^2 + O(u_{_P}^3)
\label{lam1mtay}
\eeq
\beq
\lambda_{3a} = 1 - u_{_P}^3 + O(u_{_P}^4)
\label{lam3atay}
\eeq
\beq
\lambda_{3b} = -u_{_P} -\frac{1}{2}u_{_P}^2 + O(u_{_P}^3)
\label{lam3btay}
\eeq
\beq
\lambda_{3c} = u_{_P} - \frac{1}{2}u_{_P}^2 + O(u_{_P}^3)
\label{lam3ctay}
\eeq

\begin{figure}
\epsfxsize=3.5in
\epsffile{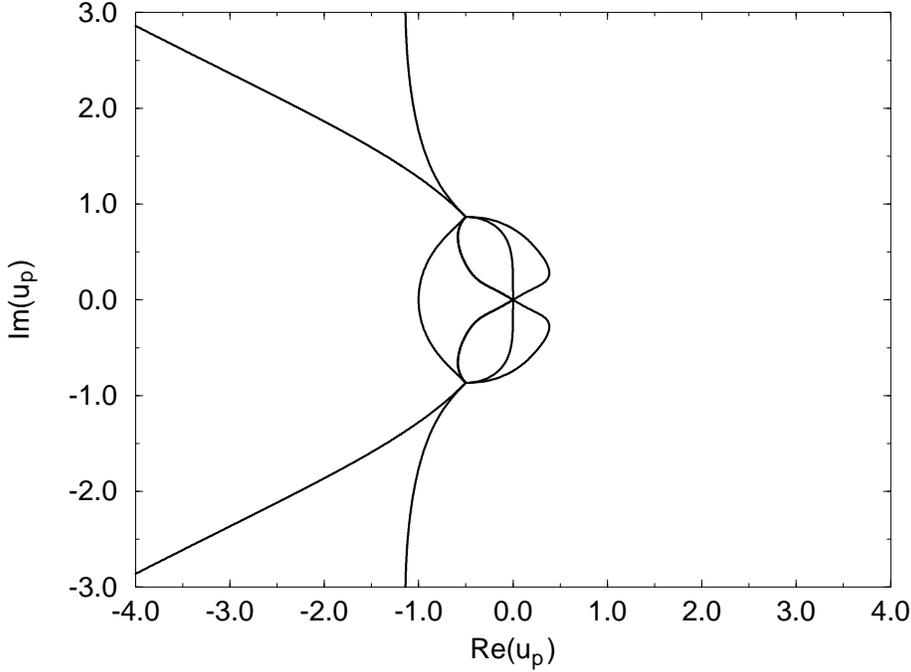}
\caption{Complex-temperature phase diagram of the 1D $q=3$ Potts model in the 
$u_{_P}$ plane for $r=1$.}
\label{figpottsr1}
\end{figure}

In the absence of any NNN coupling, 
eq. (\ref{ucircle}) shows that ${\cal B}$ would be a circle of radius 1 
centered at $u_{_P}=-1$.  In Fig. \ref{figpottsr1} we show the 
complex-temperature phase diagram for $r=1$.  We find
that the presence of the NNN interaction has a strong effect on this
diagram. The CT phase boundary is much more complicated than just the unit
circle centered at $u_{_P}=-1$.  Rather than just two regions, as in the model
with only NN spin-spin interactions, the complex-temperature phase diagram
consists of nine phases.  Three of these are 
(a) the region containing the positive real $u_{_P}$ axis, where
$\lambda_{1+}$ is dominant; (b) the region including the interval 
$-1 < u_{_P} < 0$, in which $\lambda_{3a}$ is dominant; and (c) the region
including the rest of the negative real axis, $-\infty < u_{_P} < -1$, where
$\lambda_0$ is dominant.  The remaining six are comprised of three complex
conjugate pairs.  Starting from the northeast quadrant and moving to the
northwest quadrant, the members of these pairs with $Im(u_{_P}) > 0$ are (d) a
region with a wedge contiguous to the origin, where $\lambda_{3a}$ is 
dominant; (e) a second region contiguous to the origin, where $\lambda_{1+}$ is
dominant, and (f) at the same angle, $\simeq 2\pi/3$, but farther out from 
the origin, a region where, depending on how the cuts linking the branch points
of the cube roots are chosen, $\lambda_{3b}$ or $\lambda_{3c}$ 
is dominant; the others are then the complex 
conjugates of these.  The complex-temperature boundary ${\cal B}$ contains a
multiple point at $u_{_P}=0$, where six curves come together with three
separate tangents (hence index 3), and two complex conjugate multiple points at
$(u_{_P})_m=e^{2\pi i/3}$ and $(u_{_P})_m^*$, where six curves meet in a
tacnode, with three different tangents (see Ref. \cite{cmo} for a discussion of
tacnodes on CT boundaries ${\cal B}$).  We note that the complex conjugate
boundary curves 
separating phases (a) and (f) and (a) and (f)$^*$, respectively, eventually
head outward in northeast and southeast directions at larger distance 
$|u_{_P}|$ from the origin. 

\eject

\section{Conclusions}

   In this work we have continued our exploration of the dependence of 
complex-temperature phase diagrams on details of the Hamiltonian, focussing 
on the effect of a next-nearest-neighbor spin-spin coupling in a simple 
exactly solvable model, the 1D Ising model with nearest- and 
next-nearest-neighbor spin-spin couplings.  Even for the range of values of 
$r=J_{nnn}/J_{nn}$ where the NNN coupling is an irrelevant perturbation to 
the Hamiltonian at the $T=0$ critical point, we have shown that 
it has a considerable effect on the complex-temperature phase diagram. 
We have also presented some corresponding findings for the 1D $q=3$ Potts 
model.  Our 
results further emphasize that, while complex-temperature phase diagrams and 
singularities give a deeper insight into the behavior of statistical 
mechanical models, they depend on details of the Hamiltonian, in contrast to
the usual universality observed at physical critical points. 

This research was supported in part by the NSF grant PHY-93-09888.

\vfill
\eject
\end{document}